\documentclass[useAMS,usenatbib]{mnras}

\usepackage{graphicx}
\usepackage{epstopdf}
\usepackage{amsmath,esint}

\title[Striking Confluence Between Theory and Observations of HMXB Pulsars]{A Striking Confluence Between Theory and Observations of High-Mass X-ray Binary Pulsars}

\author[D. M. Christodoulou et al.]{D. M. Christodoulou,$^{1,2}$\thanks{E-mail: dimitris\_christodoulou@uml.edu} 
S. G. T. Laycock,$^{1,3}$\thanks{E-mail: silas\_laycock@uml.edu}
and D. Kazanas$^{4}$\thanks{E-mail: demos.kazanas@nasa.gov}
\\
$^{1}$Lowell Center for Space Science and Technology, University of Massachusetts Lowell, Lowell, MA, 01854, USA\\
$^{2}$Department of Mathematical Sciences, University of Massachusetts Lowell, 
Lowell, MA, 01854, USA\\
$^{3}$Department of Physics \& Applied Physics, University of Massachusetts Lowell, Lowell, MA, 01854, USA\\
$^{4}$NASA Goddard Space Flight Center, Laboratory for High-Energy Astrophysics, Code 663, Greenbelt, MD 20771, USA
}

\begin{document}

\def\gsim{\mathrel{\raise.5ex\hbox{$>$}\mkern-14mu
                \lower0.6ex\hbox{$\sim$}}}

\def\lsim{\mathrel{\raise.3ex\hbox{$<$}\mkern-14mu
               \lower0.6ex\hbox{$\sim$}}}

\pagerange{\pageref{firstpage}--\pageref{lastpage}} \pubyear{2017}

\maketitle

\label{firstpage}

\begin{abstract}
We analyse the most powerful X-ray outbursts from neutron stars in ten Magellanic high-mass X-ray binaries and three pulsating ultraluminous X-ray sources. Most of the outbursts rise to $L_{max}$ which is about the level of the Eddington luminosity, while the rest and more powerful outbursts also appear to recognize that limit when their emissions are assumed to be anisotropic and beamed toward our direction. We use the measurements of pulsar spin periods $P_S$ and their derivatives $\dot{P_S}$ to calculate the X-ray luminosities $L_p$ in their faintest accreting (``propeller'') states. In four cases with unknown $\dot{P_S}$, we use the lowest observed X-ray luminosities, which only adds to the heterogeneity of the sample. Then we calculate the ratios $L_p/L_{max}$ and we obtain an outstanding confluence of theory and observations from which we conclude that work done on both fronts is accurate and the results are trustworthy: sources known to reside on the lowest Magellanic propeller line are all located on/near that line, whereas  other sources jump higher and reach higher-lying propeller lines. These jumps can be interpreted in only one way, higher-lying pulsars have stronger surface magnetic fields in agreement with empirical results in which $\dot{P_S}$ and $L_p$ values were not used.
\end{abstract}




\section{Introduction}\label{intro}

We are in the process of collecting and analysing X-ray observations of Magellanic high-mass X-ray binaries (HMXBs) obtained by {\it Chandra}, {\it XMM-Newton}, and {\it RXTE} over the past 15-20 years \citep{yang17,cap17,chr17a,chr16,chr17c}. The data that we have compiled from the published literature and the {\it XMM-Newton} archive show that a fair number of Magellanic pulsars ($\sim$10) were caught jumping up to about the Eddington limit $L_{Edd}$ during their brightest type I and type II \citep{coe10} outbursts \citep[Figures 2 and 4 in][]{chr16}. This is an indication that the power output from these sources accreting from discs is somehow limited to the Eddington rate despite the fact that this limiting luminosity was derived for the case of spherical wind accretion. Hence, we hypothesized that ultraluminous X-ray (ULX) sources harboring neutron stars (NSs) may also recognize the Eddington limit and in \cite{chr17b} we studied their properties in their minimum accreting (``propeller'') states \citep{ill75} assuming that their X-ray emissions are strongly anisotropic \citep[an assumption supported by the recent results of][]{dau17,kaw16}. We found that the propeller states of the three NS ULX sources are quite similar to the lowest propeller states of Magellanic Be/X-ray pulsars, implying that all of these sources share common physical properties and evolutions, except for the amount of anisotropy in their X-ray emissions.

\begin{table*}
\caption{NS ULX and HMXB Sources: Observations and Calculated Results}
\label{table1}
\begin{tabular}{cllcclclcl}
\hline
 & & & & & Observed & \multicolumn{2}{c}{Observed}   & \multicolumn{2}{c}{Calculated~~~~~~~}   \\
No. & Source & $P_S$ & $\dot{P_S}$ & $\dot{P_S}~\pm$ Error & $L_{max}$ & $L_{po}$  & $\ell_{po}$ & $L_{pc}$ & $\ell_{pc}$  \\
 & & (s) & (s s$^{-1}$) & (s s$^{-1}$) & (erg~s$^{-1}$) & (erg~s$^{-1}$) & (\%) & (erg~s$^{-1}$) & (\%)  \\
\hline
1  & NGC7793 P13 & 0.417 & $-3.5\times 10^{-11}$ & $3.0\times 10^{-14}$ & $7.0\times 10^{39}$ &  & & $5.0\times 10^{38}$ & 7.1   \\
2  & SMC X-1 & 0.72 &  & & $4.6\times 10^{37}$ & $1.2\times 10^{37}$ & 26 & &  \\
3  & NGC5907 ULX-1 & 1.137 & $-8.1\times 10^{-10}$ & $1.0\times 10^{-11}$& $1.0\times 10^{41}$ & & & $1.1\times 10^{39}$ & 1.1   \\
4  & M82 X-2 & 1.3725 & $-2.0\times 10^{-10}$ & $7.0\times 10^{-12}$& $2.0\times 10^{40}$ & & & $1.8\times 10^{38}$ & 0.90 \\
5  & SMC X-2 & 2.372 & $-1.4\times 10^{-10}$ & $9.3\times 10^{-11}$ & $4.0\times 10^{38}$ &  & & $3.4\times 10^{37}$ & 8.5  \\
6  & LXP4.10 & 4.0635 &  & & $3.7\times 10^{37}$ & $7.9\times 10^{34}$ & 0.21 & &  \\
7  & SXP4.78 & 4.782 & $-2.3\times 10^{-11}$ & $1.2\times 10^{-11}$ & $6.4\times 10^{37}$ &  & & $1.1\times 10^{36}$ & 1.7  \\
8  & SXP5.05 & 5.05 & $-6.9\times 10^{-11}$ & $1.4\times 10^{-10}$& $5.2\times 10^{37}$ &  & & $2.9\times 10^{36}$ & 5.6  \\
9  & SXP6.85 & 6.85 & $-1.2\times 10^{-11}$ & $2.3\times 10^{-11}$ & $2.4\times 10^{37}$ & &  & $2.4\times 10^{35}$ & 1.0  \\
10  & SMC X-3$^{(a)}$ & 7.78 & $-7.4\times 10^{-10}$ & $8.0\times 10^{-11}$ & $1.2\times 10^{39}$ &  & & $1.1\times 10^{37}$ & 0.92 \\
11  & LXP8.04 & 8.035 &  & & $8.0\times 10^{38}$$^{(b)}$ & $5.0\times 10^{34}$  & 0.0063 & &  \\
12  & LMC X-4$^{(c)}$ & 13.50 & $-6.8\times 10^{-11}$ & $6.3\times 10^{-13}$ & $4.0\times 10^{39}$ & &  & $2.9\times 10^{35}$ & 0.0073  \\
13  & SXP15.3 & 15.239 &  & & $8.6\times 10^{37}$ & $6.8\times 10^{33}$  & 0.0079 &  &  \\
\hline
\end{tabular}
\\
\smallskip
(a)~Data from \cite{tow17} and \cite{weng17}. \\ 
(b)~Here we correct a typographical error in Table 1 of \cite{chr16}. \\
(c)~Data from \cite{sht18} and \cite{lev00}.
\end{table*} 

Thus it is reasonable to study a combined sample of the above sources and, in particular, their state transitions between the highest ($L_{max}\gsim L_{Edd}$) and lowest ($L_p$) accreting states. This is now viable because we have measurements of the spin periods $P_S$ and their time derivatives $\dot{P_S}$ for Magellanic HMXB pulsars \citep{yang17} and for the NS ULX sources 
\citep{bac14,fur16,isr17a,isr17b}.

The sample that we have compiled for this analysis is shown in Table~\ref{table1}. It consists of ten Magellanic sources and three NS ULX sources, all of which have been observed to emit near or above $L_{Edd}$ during their most powerful X-ray outbursts. In \S~\ref{data}, we calculate their propeller luminosities $L_p$ and surface dipolar magnetic fields $B$ based on the measured values of $P_S$ and $\dot{P_S}$ and for canonical pulsar parameters (a mass of 1.4$M_\odot$ and a radius of 10~km). We find that this is a homogeneous data set in which the pulsars share similar states and state transitions and they can be distinguished only by their $B$ values. But the $B$ values have been determined independently without using $\dot{P_S}$ or $L_p$ values \citep{chr17c}, and the agreement between the two determinations is outstanding. This, in turn, demonstrates that both the measurements and the theoretical calculations are accurate and trustworthy. In \S~\ref{sum}, we discuss these results.

\begin{figure}
\begin{center}
    \leavevmode
      \includegraphics[trim=0 0cm 0 0.1cm, clip, angle=0,width=9 cm]{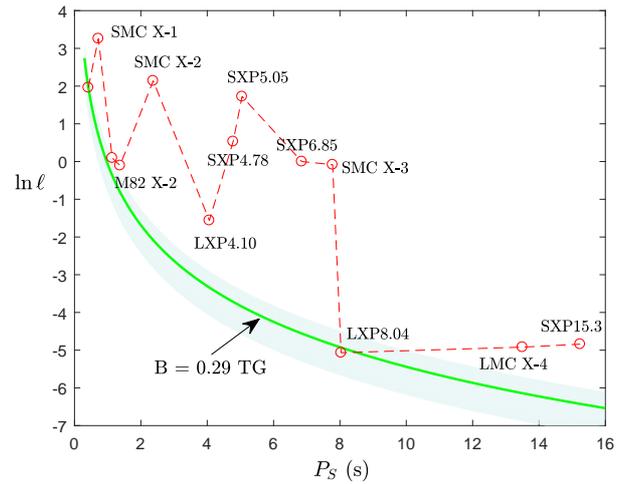}
\caption{The propeller luminosity $L_p$ is expressed as a percentage of the maximum observed luminosity $L_{max}$ and is plotted versus pulsar spin period $P_S$ for the X-ray sources listed in Table~\ref{table1}. The solid line is the lowest propeller line with characteristic surface dipolar magnetic field $B=0.29$~TG found in Magellanic HMXB pulsars and expressed as a percentage of the Eddington rate for the canonical NS mass of 1.4$M_\odot$. The shaded area illustrates the error bar in the lowest propeller line ($B=0.20$-0.36~TG). Pulsars with significantly stronger magnetic fields jump high above the lowest propeller line in order to reach their own higher-lying propeller lines.
\label{fig1}}
  \end{center}
\end{figure}

\section{X-ray Data and Analysis}\label{data}

Table~\ref{table1} summarises the observations and the results of our $L_p$ calculations. The measurements of NS ULX sources were obtained from 
\cite{bac14}, \cite{fur16}, and \cite{isr17a,isr17b}; the measurements of Magellanic HMXB sources were obtained from \cite{yang17}, \cite{chr16}, and the latest observations listed in the notes to the table. The $L_{max}$ column shows the observed X-ray luminosities that have reached near or above $L_{Edd}=1.8\times 10^{38}$~erg~s$^{-1}$ (for a mass of 1.4$M_\odot$). The $\dot{P_S}$ column shows the existing measurements of $\dot{P_S}$ and the next column shows the published error bars. For the ULX sources, SMC X-3, and LMC X-4, the $\dot{P_S}$ values were obtained during outbursts, whereas for the remaining sources they are long-term averages obtained over a period of 15 years. The $L_{pc}$ column shows the calculated values of $L_p$ (see below). For those pulsars with no $\dot{P_S}$ measurements, we adopted the observed minimum X-ray luminosities $L_{po}$ determined in \cite{chr16} to lie near and above the lowest Magellanic propeller line with magnetic field $B=0.29$~TG.

\begin{table}
\caption{ Luminosity Jumps and NS Dipolar Magnetic Fields}
\label{table2}
\begin{tabular}{clll}
\hline
No. & Source & $\ell/\ell_1\,^{(a)}$ & $B$~(TG) \\
\hline
\multicolumn{4}{c}{Lowest Propeller Line, $B=(0.28\pm0.08)$~TG} \\
\hline
1  & NGC7793 P13    	& 1    & 0.29 \ $\pm 0.0001$ \\
3  & NGC5907 ULX-1	& 1.5 & 0.37 \ $\pm 0.002$ \\
4  & M82 X-2            	& 2    & 0.41 \ $\pm 0.007$ \\
6  & LXP4.10  & 6$^{(b)}$ & 0.32 \\
11  & LXP8.04 			& 1   & 0.27 \\
\hline
\multicolumn{4}{c}{Second Propeller Line, $B=(0.55\pm0.11)$~TG} \\
\hline
2  & SMC X-1               & 13 & 0.53 \\
12 & LMC X-4               & 3.5 & 0.53 \ $\pm 0.003$ \\
13  & SXP15.3 			   & 5  & 0.44 \\
\hline
\multicolumn{4}{c}{Third Propeller Line, $B=(1.3\pm0.35)$~TG} \\
\hline
5  & SMC X-2$^{(c)}$ & 68 & 2.4 \ $+0.70~/$ $-1.0$ \\
\hline
\multicolumn{4}{c}{Fourth Propeller Line, $B=(2.9\pm0.20)$~TG} \\
\hline
8  & SXP5.05 			& 262 & 2.5 \ $+1.9~/ \cdots$ \\
10 & SMC X-3            & 118 & 3.1 \ $\pm 0.17$ \\
\hline
\multicolumn{4}{c}{Highest Propeller Line, $B=(8.2\pm2.7)$~TG} \\
\hline
7  & SXP4.78     		& 70 & 1.5 \ $+0.34~/$ $-0.45$ \\
9  & SXP6.85 			& 95 & 1.0 \ $+0.73~/ \cdots$ \\
\hline
\end{tabular}
\\
\smallskip
(a)~Jump factors above the lowest propeller line $\ell_1$. \\
(b)~See footnote \ref{ft1}. \\
(c)~\cite{jai16} determined that $B=2.3$ TG from a cyclotron absorption line detected by {\it NuSTAR} and {\it Swift/XRT}.
\end{table} 

The dimensionless ratios are defined by $\ell_{po}\equiv L_{po}/L_{max}$ and $\ell_{pc}\equiv L_{pc}/L_{max}$ and they are expressed in percentage form. For all sources with measured $\dot{P_S}$ values, the $L_{pc}$ values were calculated from the equation \citep{chr1416,chr17b}
\begin{equation}
L_{pc} = \frac{\eta}{2}\left(2\pi I_* |\dot{P_S}|\right)\left( \frac{2\pi}{P_S^7}\frac{G M}{R_*^3} \right)^{1/3} \, ,
\label{lmin2}
\end{equation}
where $\eta/2$ is the efficiency of converting accretion power to X-rays ($\eta$ taken here to be 0.5), $I_*$ is the NS moment of inertia, $G$ is the gravitational constant, and $M$ and $R_*$ are the canonical values of the NS mass and radius, respectively. The introduction of $\eta/2$ also implies that we assume that minimum accretion takes place at a reduced torque compared to the observed maximum values during outbursts. The overall value of $\eta/2$ is uncertain; in general, we choose $\eta/2=0.25$ for NSs and $\eta/2=0.5$ for black holes, based in part on the work of \cite{kor06}. Our choice of $\eta/2=0.25$ for NSs is also supported by the observations of M82 X-2: assuming isotropic emission, eq.~(\ref{lmin2}) gives $L_{pc}=1.8\times 10^{38}$~erg~s$^{-1}$ and this value agrees well with the value determined by \cite{tsy16}. From 15 years of {\it Chandra} observations and assuming isotropic emission, \cite{tsy16} found that M82 X-2 has repeatedly switched between its high state and a low state with $L_{po}=1.7\times 10^{38}$~erg~s$^{-1}$ \citep[see also][who obtained the same value as an upper limit in quiescence]{dal16}.

In this seemingly heterogeneous data set, we finally define the dimensionless ratio $\ell\equiv L_p/L_{max}$ with no distinction between $L_{pc}$ and $L_{po}$ or $\ell_{pc}$ and $\ell_{po}$. The reason for using this ratio rather than propeller values is that uncertainties about the distances and beaming characteristics of these sources, especially the ULX sources, are removed. The ratio $\ell$ for observed quantities is equal to the ratio of the observed X-ray fluxes. 

We plot in Fig.~\ref{fig1} the natural logarithms of $\ell$ versus $P_S$. The individual points are connected by a dashed line. We also plot for reference the lowest Magellanic propeller line scaled to $L_{Edd}$ (solid line). Its $\ell =\ell_1$ ratio is defined for canonical pulsar parameters ($M=1.4 M_\odot$ and $R_*=10$~km) by the equation \citep{ste86,chr17b}
\begin{equation}
\ell_1\equiv \frac{L_{p1}}{L_{Edd}} =  9.34\times 10^{-3}\left(\frac{P_S}{1~{\rm s}}\right)^{-7/3} \, ,
\label{stella}
\end{equation}
where $L_{p1}$ is given by eq.~(5) of \cite{chr17b} with magnetic moment $\mu\equiv B R_*^3$ and $B=0.29$~TG \citep{chr17c}.

At first glance, the result depicted in Fig.~\ref{fig1} is surprising. The ratio $\ell$ is expected to decrease steadily with spin period, just as the theoretical (solid) line does. Instead, the pulsars show a sawtooth distribution of $\ell$ vs. $P_S$. Some of them appear to stay near the theoretical propeller line (the ULX sources, LXP4.10,\footnote{LXP4.10 did not have a type II outburst \citep{sch95} and its $L_{max}$ is considerably lower than $L_{Edd}$, which pushes the point higher in Fig.~\ref{fig1}. Nevertheless it belongs to the lowest propeller line, as its $B$ value in Table~\ref{table2} indicates.\label{ft1}} and LXP8.04), but others jump consistently higher above that line. The jumps are pronounced and they cannot be explained as a result from observational errors, uncertainties surrounding $\dot{P_S}$ and $\eta$ and their influence on eq.~(\ref{lmin2}), or concerns about the validity of the equations themselves.

The characteristic behaviour of $\ell$ values seen in Fig.~\ref{fig1} and Table~\ref{table1} can be explained as follows: the past observations of low-power states of Magellanic pulsars indicate that not all of them reach the lowest propeller line with $B=0.29$~TG \citep{chr17c}. In fact, we have discovered five different propeller lines each characterized by a progressively higher value of the magnetic field. The sources that jump in Fig.~\ref{fig1} have all been predicted to belong to higher propeller lines. Thus, the surprising element in this analysis is not the sawtooth behaviour, but the agreement between the present results and the empirical conclusions of Christodoulou et al. (see Table~1 in \cite{chr17c}) who did not use $\dot{P_S}$ or $L_{pc}$ in their determination of multiple propeller lines. 

In Table~\ref{table2}, we quantify the degree of agreement in the determination of magnetic fields between this work and the results in \cite{chr17b} and \cite{chr17c}. The headings show the five empirical propeller lines from \cite{chr17c} and the tight ranges of the segregated magnetic fields of Magellanic pulsars that defined them. These magnetic fields were calculated from the equation for minimum accretion of \cite{ste86} using only the lowest observed X-ray luminosities at progressively higher luminosity levels. The body of Table~\ref{table2} lists our new determinations of the magnetic fields using data from Table~\ref{table1}, most of which depend on the measured values of $\dot{P_S}$. The sources have been regrouped according to the predictions of \cite{chr17c} for Magellanic HMXB sources. Beaming has been taken into account for the sources whose brightest outbursts have registered significantly higher than the Eddington limit \citep[the NS ULX sources, SMC X-2, SMC X-3, LXP8.04, and LMC X-4; for details, see][]{chr17b}. The agreement on the lower four propeller lines is precise. The difference for SMC X-2 may be explained due to the large error bar in the determination of $\dot{P_S}$ ($\pm$66\% in Table~\ref{table1}); but see also note (c) in Table~\ref{table2} for a counterargument that places SMC X-2 on the fourth propeller line.\footnote{In \cite{chr17c}, SMC X-2 was placed on the third propeller line based on a single very faint {\it ASCA} detection \citep[$L_{po}=3.9\times 10^{36}$ erg~s$^{-1}$;][]{yok03}.} 

The results differ substantially only in the fifth (highest) propeller line and the difference cannot be explained due to the error bars in the determination of $\dot{P_S}$. But this is the only line that is highly uncertain in the classification of \cite{chr17c}, and the two sources involved could very well belong to the third propeller line since their jump factors are only $\ell/\ell_1=70$ and 95 (Table~\ref{table2}); additionally, only one observation was available for each source in our database and these single observations played a big role in defining the highest (and uncertain) propeller line.

The outstanding agreement between results in the lower four propeller lines supports the confluence between theory and observations in our disparate sample of sources, as well as the hypothesis that ULX NS sources are strongly beamed in the direction of the observer. Despite their apparently extreme power output, these NSs appear to be quite similar to Magellanic pulsars and their surface dipolar magnetic fields (0.3-0.4~TG) take some of the lowest values ever found in studies of HMXBs \citep[see also Fig.~1 in][]{chr17b}.

\section{Discussion}\label{sum}

We have studied a sample of NS ULX and Magellanic NS HMXB sources that are known to radiate near the Eddington rate or higher during their brightest outbursts. This sample consists of three NS ULX sources and ten HMXB sources as indicated in Table~\ref{table1} and their X-ray properties and measurements have been summarised and analysed, among all other Magellanic Be/X-ray sources, by \cite{yang17} and in \cite{chr17b}, \cite{chr16}, and \cite{chr17c}.
 
This work was motivated by the realization that we have two independent ways of determining the surface dipolar magnetic fields $B$ of these NSs: one method relies on the observed X-ray luminosity $L_{po}$ at the state of minimum accretion (the propeller state) and the assumption that the magnetospheric radius is equal to the corotation radius \citep{ill75,ste86}; the other method uses the observed value of $\dot{P_S}$ (which is unjustly believed to be highly uncertain) and canonical pulsar parameters to obtain direct estimates of both $L_{pc}$ and $B$ from standard accretion theory \citep{fra02,chr1416,chr17b}. These two methods showed good agreement when they were applied to M82 X-2 \citep{chr17b}, but that result has gone largely unnoticed. 

Thus we set out to investigate whether the agreement between the two methods extends to more HMXB pulsars, specifically to those who have exhibited the most powerful outbursts during their 15-20 year monitoring by X-ray observatories. Our results for the $L_p/L_{max}$ ratios are shown in Fig.~\ref{fig1} and the results for the magnetic field  $B$ values are summarised in Table~\ref{table2}. We have found oustanding agreement between the two deteminations of $B$ in conjunction with past predictions for Magellanic HMXB pulsars \citep[Table~1 in][]{chr17c}. This agreement indicates that there is a strong confluence between theory and observations of HMXB pulsars, which is very encouraging for future investigations and investments in the field. Furthermore, the recent results of \cite{cam17} show that the lowest propeller line determined from theory \citep[$L_{pc}$ vs. $P_S$;][]{ste86} and independently from Magellanic pulsars \citep[$L_{po}$ vs. $P_S$;][]{chr16} is applicable to all rotating magnetised stars covering eight orders of magnitude in spin period and ten orders of magnitude in magnetic moment.

We close with an observation pertaining to the data listed in Table~\ref{table1} that leads to an open issue: the results of \cite{yang17} and \cite{chr17a} on accretion discs in Magellanic HMXBs indicate that about half of them are spinning up and the other half are spinning down. On the other hand, Table~\ref{table1} shows that all nine sources in our sample with measured $\dot{P_S}$ values are spinning up. It appears then that major type II outbursts \citep{coe10} do not occur when the accretion discs or inflowing streams orbit in the opposite direction relative to the spins of the NSs. There is no concrete explanation for this phenomenon at this time, although it certainly appears that the spinning down NSs during retrograde accretion ($\dot{P_S} > 0$) produce lower power output emanating from their accretion columns.

\section*{Acknowledgments}
DMC and SGTL were supported by NASA grant NNX14-AF77G.
DK was supported by a NASA ADAP grant.

\label{lastpage}

\end{document}